\documentclass[prl,twocolumn,longbibliography,superscriptaddress]{revtex4-1}
\usepackage{amsmath}
\usepackage{amssymb}
\usepackage{amsthm}
\usepackage{amsfonts}
\usepackage{listings}
\usepackage{enumerate}
\usepackage{latexsym}
\usepackage{psfrag}
\usepackage{bm}
\usepackage[all]{xy}
\usepackage{graphicx}
\usepackage{subfigure}
\usepackage[pdftex]{hyperref}
\usepackage{color}
\usepackage{mathtools}
\usepackage{ gensymb }
\usepackage{lipsum}

\begin{document}

\title{Topological Hall effect in bulk ferromagnet Cr$_2$Te$_3$  embedded with  black-phosphorus-like bismuth nanosheets}

\author{Liang Zhou}
\thanks{These three authors contribute to this work equally.}
\affiliation{Institute for Quantum Science and Engineering, and Department of Physics, Southern University of Science and Technology, Shenzhen 518055, China}
\author{Junshu Chen}
\thanks{These three authors contribute to this work equally.}
\affiliation{Institute for Quantum Science and Engineering, and Department of Physics, Southern University of Science and Technology, Shenzhen 518055, China}
\author{Xiaobin Chen}
\thanks{These three authors contribute to this work equally.}
\affiliation{School of Science, Harbin Institute of Technology, Shenzhen 518055, China}
\affiliation{Institute for Quantum Science and Engineering, and Department of Physics, Southern University of Science and Technology, Shenzhen 518055, China}
\thanks{These three authors contribute to this work equally.}
\author{Bin Xi}
\affiliation{College of Physics Science and Technology, Yangzhou University, Yangzhou 225002, China}
\author{Yang Qiu}
\affiliation{Materials Characterization and Preparation Center, Southern University of Science and Technology, Shenzhen 518055, China}
\author{Junwei Zhang}
\affiliation{King Abdullah University of Science and Technology (KAUST), Physical Science and Engineering,
Thuwal 23955-6900, Saudi Arabia}

\author{Linjing Wang}
\affiliation{Institute for Quantum Science and Engineering, and Department of Physics, Southern University of Science and Technology, Shenzhen 518055, China}
\author{Runnan Zhang}
\affiliation{Institute for Quantum Science and Engineering, and Department of Physics, Southern University of Science and Technology, Shenzhen 518055, China}
\author{Bicong Ye}
\affiliation{Institute for Quantum Science and Engineering, and Department of Physics, Southern University of Science and Technology, Shenzhen 518055, China}
\author{Pingbo Chen}
\affiliation{Institute for Quantum Science and Engineering, and Department of Physics, Southern University of Science and Technology, Shenzhen 518055, China}
\author{Xixiang Zhang}
\affiliation{King Abdullah University of Science and Technology (KAUST), Physical Science and Engineering,
Thuwal 23955-6900, Saudi Arabia}
\author{Guoping Guo}
\affiliation{Key Laboratory of Quantum Information, CAS, University of Science and Technology
of China, Hefei 230026, China}

\author{Dapeng Yu}
\affiliation{Institute for Quantum Science and Engineering, and Department of Physics, Southern University of Science and Technology, Shenzhen 518055, China}
\author{Jia-Wei Mei}
\email{meijw@sustech.edu.cn}
\affiliation{Institute for Quantum Science and Engineering, and Department of Physics, Southern University of Science and Technology, Shenzhen 518055, China}
\author{Fei Ye}
\email{yef@sustech.edu.cn}
\affiliation{Institute for Quantum Science and Engineering, and Department of Physics, Southern University of Science and Technology, Shenzhen 518055, China}
\author{Gan Wang}
\email{wangg@sustech.edu.cn}
\affiliation{Institute for Quantum Science and Engineering, and Department of Physics, Southern University of Science and Technology, Shenzhen 518055, China}
\affiliation{Shenzhen Key Laboratory of Quantum Science and Engineering, Shenzhen 518055, China.}
\author{Hongtao He}
\email{heht@sustech.edu.cn}
\affiliation{Institute for Quantum Science and Engineering, and Department of Physics, Southern University of Science and Technology, Shenzhen 518055, China}

\date{\today}

\begin{abstract}
We implement the molecular beam epitaxy method to embed the
black-phosphorus-like  bismuth nanosheets into the bulk ferromagnet
Cr$_2$Te$_3$. As a typical surfactant, bismuth lowers the surface tensions and mediates the
layer-by-layer growth of Cr$_2$Te$_3$. Meanwhile, the bismuth atoms
precipitate into black-phosphorus-like nanosheets with the lateral size of several
tens of nanometers. In Cr$_2$Te$_3$ embedded with Bi-nanosheets, we observe simultaneously a large topological Hall effect together
with the magnetic susceptibility plateau and magnetoresistivity
anomaly.  As a control experiment, none of these signals is observed in the
pristine Cr$_2$Te$_3$ samples. Therefore, the Bi-nanosheets serve as seeds of
topological Hall effect induced by non-coplanar magnetic textures planted into Cr$_2$Te$_3$.  Our experiments demonstrate a new method to
generates a large topological Hall effect by planting strong spin-orbit
couplings into the traditional ferromagnet, which may have potential
applications in spintronics.
\end{abstract}

\maketitle

\emph{Introduction --}
When it flows in the presence of a static magnetic field in the
perpendicular direction, the electron is deflected by the Lorentz
force and gives rise to the Hall effects identified by a transverse
resistivity~\cite{Hall1879}.  The external magnetic field is not mandatory, however, `Berry curvature' is essential  to
produce a Hall effect~\cite{Nagaosa2010}. The spin-orbit coupling in the magnetic
system generates the anomalous Hall effect (AHE) associated
with the Berry curvature in momentum space~\cite{Nagaosa2010}. The non-coplanar
magnetic texture could have a chiral real space configuration of the local
moments, which have unique geometrical or topological
properties in real space leading to the topological Hall effect (THE)~\cite{Nagaosa2012,Nagaosa2012a,Nagaosa2013,Ye1999,Bruno2004,Neubauer2009,Matsuno2016,Yasuda2016,Zhao2017,Liu2017,Lee2009,Kanazawa2011,Huang2012}. 
THE is usually  observed in mainly three families of
materials, such as the chiral magnetic B20
compounds (MnSi, FeGe, Fe$_{1-x}$Co$_x$Si, etc.) with broken inversion
symmetry~\cite{Muehlbauer2009,Neubauer2009,Grigoriev2009a,Wilhelm2011,Tokunaga2015, Karube2016},
the metallic 
ferromagnet/paramagnet bilayers with artificial Dzyaloshinsky-Moriya (DM) interactions at the
interface~\cite{Bode2007,Matsuno2016,Yasuda2016,Zhao2017},
and magnetically doped topological
insulators~\cite{Ye2010,Liu2017}. The THE is a signature of the chiral magnetic texture, \textit{i.e.}
magnetic skyrmions~\cite{Skyrme1962,Bogdanov1989,Bogdanov1994,Nagaosa2013},
which may have potential applications in spintronics. 

THE originates in the entanglement of
ferromagnets and strong spin-orbit couplings. 
To expand the family of `THE materials', it is of significant interests to
efficiently insert strong spin-orbit couplings into
traditional ferromagnets, which is still challenging so far.
Among many typical ferromagnets, the study of chromium tellurides has a
long history and can be traced back to
1935~\cite{Haraldsen1935}. Depending on the stoichiometric
ratio, Cr$_{1-x}$Te has a rich structural phase diagram with Curie
temperatures ranging from 170~K to 340~K~\cite{Haraldsen1935}.  Very recently,
the Cr$_{2}$Te$_3$ thin film with a high-quality single crystalline
structure has been synthesized using the molecular beam epitaxy (MBE)
technique, which shows a strong perpendicular magnetic anisotropy and a
large magnetic moment per Cr atom about 2.8
$\mu_B$~\cite{Roy2015,Zhou2017}. If strong spin-orbit coupling elements
can be doped in an appropriate way, Cr$_2$Te$_3$ would exhibit significant THE with the onset temperature close to the
Curie temperature. A convenient candidate is bismuth, which is well
known not only for its strong spin-orbit coupling, but also as a typical
surfactant to achieve the layer-by-layer growth mode in
MBE~\cite{Copel1989,Vegt1992}. Furthermore, bismuth has
various atom-thick layer structures, such as bismuthene~\cite{Reis2017}
and the very unusual puckered-layer
structure~\cite{Nagao2004,Yaginuma2008} similar to black
phosphorous~\cite{Akhtar2017}.

\begin{figure*}[t]
\centering
\includegraphics[width=2\columnwidth]{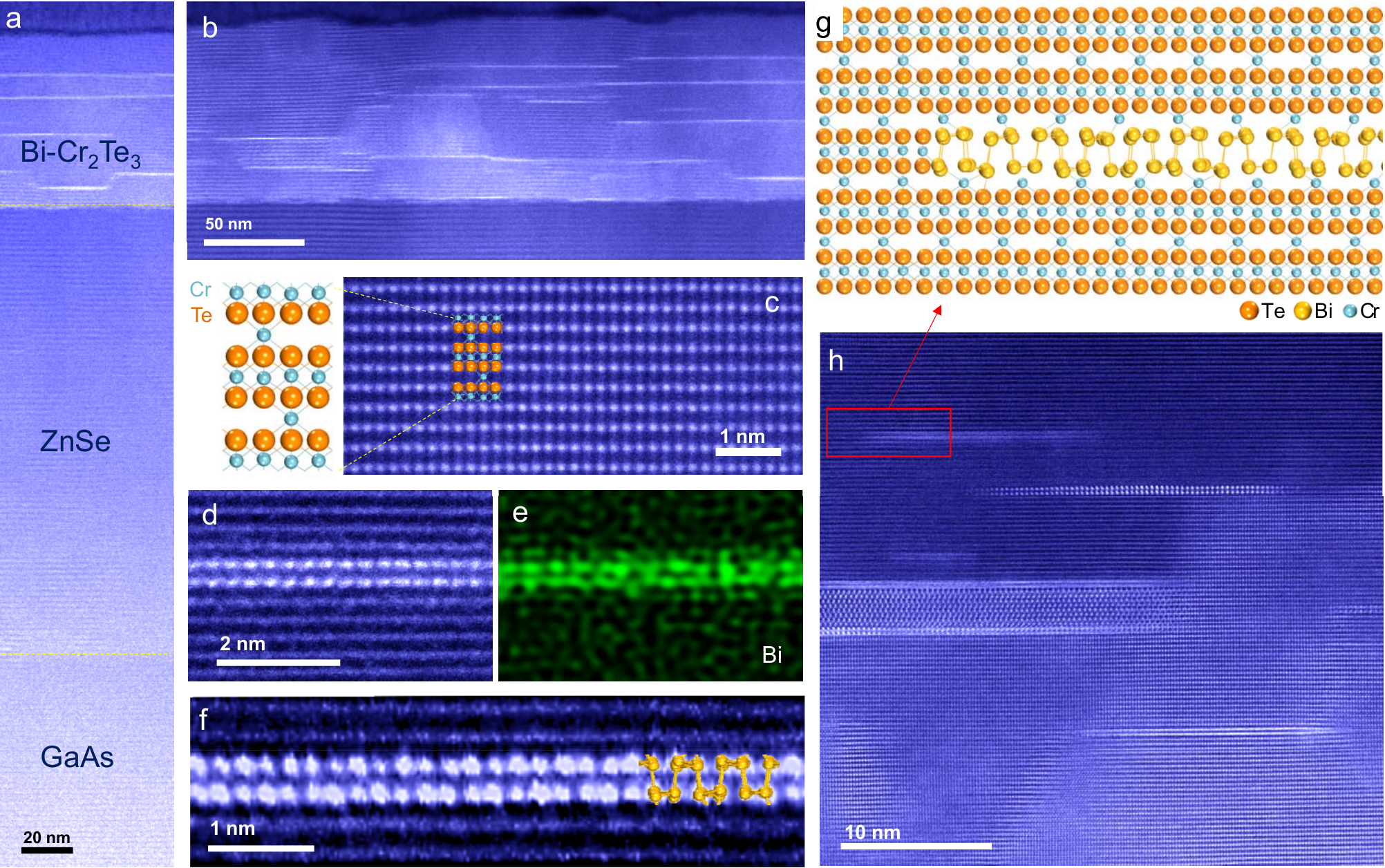}
\end{figure*}
\begin{figure*}
\caption{ HAADF STEM images  of the MBE-grown
  Bi-embedded Cr$_2$Te$_3$. (a) Cross-sectional HAADF STEM image of a sliced
  MBE-grown Cr$_2$Te$_3$ sample. The visible white filaments inside the
  Cr$_2$Te$_3$ layer are bismuth nanosheets. (b) Landscape STEM image of a
  Cr$_2$Te$_3$ sample. (c) High resolution STEM image of crystalline
  Cr$_2$Te$_3$. 
  Note that the Cr atoms are invisible in the STEM image. (d) and (e) STEM image
  and EDS mappings for the distribution of Bi element. (f) STEM image with
  enlarged scale showing a bilayer bismuth sandwiched between layers of
  Cr$_2$Te$_3$. (g) Atomic configuration (side view) of a bismuth nanosheet
  embedded Cr$_2$Te$_3$ structure. 
  (h) High resolution STEM image for several Bi nanosheets embedded in the Cr$_2$Te$_3$ layer.
}
\label{fig:figure1}
\end{figure*}

In this Letter, we report a new efficient route towards
  producing ``THE materials'' by embedding black-phosphorus-like
  bismuth nanosheets into the bulk ferromagnet Cr$_2$Te$_3$. We reveal a large, high-temperature
  topological Hall effect together with magnetic susceptibility plateau and
  magnetoresistivity anomaly.  As a control experiment, none of these
  signals is observed in the pristine Cr$_2$Te$_3$ samples, revealing the
  critical role of Bi-nanosheets with strong spin-orbit coupling. Therefore,
  the Bi-nanosheets serve as seeds of spin-orbit couplings planted into
  Cr$_2$Te$_3$ to generate THE. We also discuss the magnetic skyrmion scenario accounting for
  our experimental results.

\begin{figure*}
\centering
\includegraphics[width=2\columnwidth]{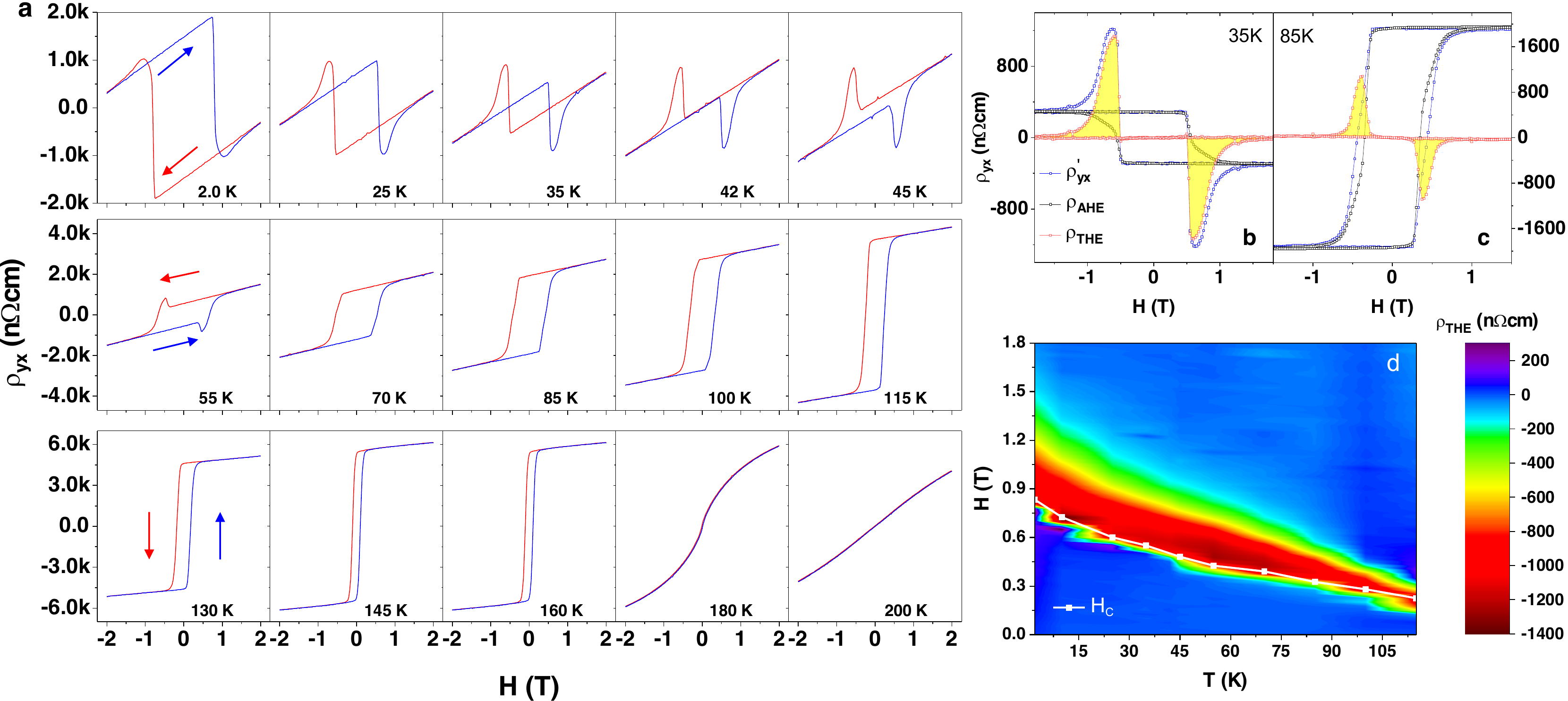}
\caption{Topological Hall Effect.
 (a) Magnetic field ($H$) dependence of Hall resistivity ($\rho_{yx}$) at various
temperatures. The red (blue) curves were measured in decreasing(increasing) the
magnetic field.  (b) and (c)
Extraction of THE contribution to Hall resistivity at 35 and 85~K. The
topological Hall resistivity ($\rho_{\text{THE}}$) is obtained by subtracting
the anomalous Hall resistivity ($\rho_{\text{AHE}}$) from $\rho'_{yx}$, where
$\rho'_{yx}=\rho_{yx}-\rho_{\text{OHE}}$ and $\rho_{\text{OHE}}$ is the ordinary
Hall resistivity.   (d)  Color map of THE in the $T$-$H$ plane. The temperature dependence of $H_c$ which is determined by the magnetic hysteresis measurement is also plotted. }
\label{fig:figure2}
\end{figure*}

\emph{Sample growth and characterizations --} 
The growth of Bi-intercalated Cr$_2$Te$_3$ thin films were performed on semi-insulating epi-ready GaAs(111)B substrates in a home-built molecular beam epitaxy system with a base vacuum of MBE better than 5$\times$10$^{-10}$ torr. Prior to the growth of Cr$_2$Te$_3$, the GaAs(111)B substrate was first deoxided at 580~$^{\circ}$C until streaky RHEED patterns appeared, followed by a deposition of 150~nm ZnSe buffer layer for smoothening the voids of GaAs induced by the deoxidation. Cr$_2$Te$_3$ thin film were grown by co-evaporating Bi (99.995\%), Cr (99.999\%), and Te (99.999\%) from Knudsen cells with a flux ratio of 2:1:10 at $T_{\text{substrate}} = 260~^{\circ}$C.
A spherical aberration corrected scanning transmission electron microscopy (STEM) were employed for performing systematic structure and chemical analyses. 

The Cr$_2$Te$_3$ thin films were grown on the GaAs(111) substrate with a buffer layer of
ZnSe, as shown in the cross-sectional high angle annular dark field (HAADF) STEM image in Fig.~\ref{fig:figure1}~(a).  During the growing process a concurrent bismuth
flux was applied together with Cr and Te fluxes. The participation of
bismuth atoms can lower the surface tensions resulting in a
surfactant-mediated layer-by-layer growth mode instead of a
three-dimensional cluster mode~\cite{Copel1989,Vegt1992}.  In
fact, we observed the extended streaky reflection high-energy electron
diffraction (RHEED) patterns during the entire growing process. The X-ray
diffraction measurement also shows that Bi-embedded samples have much better
crystalline quality than non-Bi-embedded ones~\footnote{See supplementary
  materials for more details.}. 

Figure.~\ref{fig:figure1}~(b) shows the landscape STEM image of a Cr$_2$Te$_3$ thin film embedded with Bi
nanosheets, where a single crystalline structure is found with a sharp
interface near the ZnSe buffer layer. The high-resolution STEM image of Cr$_2$Te$_3$ lattice in Fig.~\ref{fig:figure1}~(c) shows a typical NiAs-type lattice
structure of the pristine Cr$_2$Te$_3$ oriented along [001] direction.
The lattice constants of Cr$_2$Te$_3$ are determined to be
$a=b=6.81$~\AA~ and $c=12.28$~\AA.  The atom distributions of a Bi nanosheet are revealed in Fig.~\ref{fig:figure1}~(d) and (e) by the atomically resolved energy
dispersive X-ray spectroscopy (EDS) mapping, showing that the Bi atoms occupy Te sites to form a bilayer structure embedded in the Cr$_2$Te$_3$ lattice. The bismuth atoms precipitate into nanosheets with lateral size of several tens of
nanometers, corresponding to the luminescent line segments in
Fig.~\ref{fig:figure1}~(a) and (b). The atomically resolved STEM image given in
Fig.~\ref{fig:figure1}~(f) further reveals that the bismuth
nanosheet owns a bilayer structure with lattice constants
$a=4.5$~\AA~ and $c=6.5$~\AA, in a striking analogy to the
puckered-layer structure of black phosphorus~\cite{Akhtar2017}.
Fig.~\ref{fig:figure1}~(g) is the atomic configuration of a Bi-nosheet embedding
in the Cr$_2$Te$_3$ as shown in the high resolution STEM image in Fig.~\ref{fig:figure1}~(h).
The embedded bilayer bismuth exhibits a slight distortion due to the interaction
with Cr$_2$Te$_3$.

\emph{Topological Hall effect --}
Then Bi-embedded Cr$_2$Te$_3$ samples hybridize strong spin-orbit couplings and
ferromagnetism, and we observe a significant THE in the magneto-transport experiments.
Fig.~\ref{fig:figure2} is the THE result in a 60-nm-thick sample of Cr$_2$Te$_3$
embedded with Bi nanosheets. The Curie temperature ($T_c$) of the sample is determined
  to be 180~K from the
temperature dependence of resistivity and magnetization~\cite{Note1}.
Fig.~\ref{fig:figure2}~(a) shows the double-sweep measurements of the Hall resistivity
$\rho_{yx}(H)$ at various temperatures with the magnetic field $H$ along the
[001] direction. Above $T_c$, the Hall resistivity is linearly
dependent on $H$, namely, $\rho_{\text{OHE}}=R_oH$ as expected for the
ordinary Hall effect (OHE).  Since the ordinary Hall coefficient $R_o>0$
as shown in Fig.~\ref{fig:figure2}~(a), the charge carrier is of
$p$-type and the density can also be estimated to be around
$ 10^{21}$~cm$^{-3}$~\cite{Note1}, coming from the unoccupied Cr-3$d$ orbitals. Below
$T_c$, a clear hysteresis behavior is observed, indicative of the
anomalous Hall effect (AHE).  The anomalous Hall resistivity depends on
the magnetization $M$ and is generally written as
$\rho_{\text{AHE}}=R_aM$.  There are usually three independent origins
of the AHE, including the skew-scattering, side-jump and the Berry
curvature in momentum space~\cite{Nagaosa2010}, which may have opposite
contributions, thus possibly cancel with one another leading to a
vanishing anomalous Hall coefficient $R_a$~\cite{Zhou2017}. Indeed, we find the sign of
$R_a$ is reversed at $T\sim 42$~K with decreasing temperature, and no
anomalous Hall resistivity is detected right at this temperature.

\begin{figure*}[t]
\centering
\includegraphics[width=2\columnwidth]{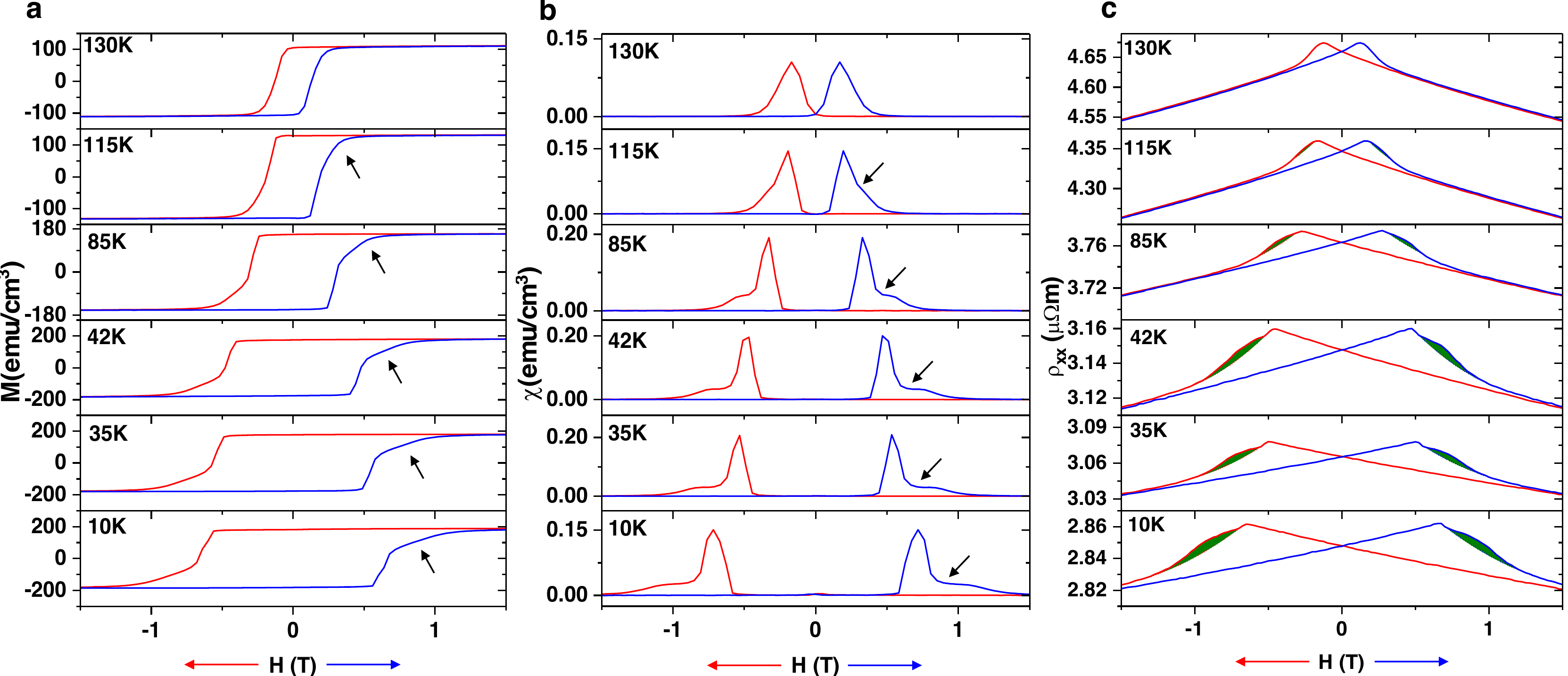}
\caption{Magnetization tail, magnetic susceptibility plateau, and magnetoresistivity anomaly.  (a) and (b) Magnetic field dependence of magnetization and the corresponding
  susceptibility, respectively, at various temperatures. The magnetic susceptibility also displays a plateau as indicated by
  an arrow.  (c) Magnetic field dependence of the longitudinal magnetoresistivity
  at various temperatures. A hump feature is clearly observed, in close
  correlation with the occurrence of THE. }
\label{fig:figure3}
\end{figure*}
The most striking
feature of the Hall resistivity curves is the emergence of abnormal
peaks in the vicinity of the coercive field ($H_c$), as shown in
Fig.~\ref{fig:figure2}~(a). These peaks show up regardless of the sign
of $R_a$, even at $T=42$~K where the AHE vanishes completely, indicating that these abnormal Hall resistivity peaks should have a
different origin from the AHE. In fact these pronounced peaks
can be attributed to the THE, which is related
to the spin chirality of magnetic skyrmions~\cite{Nagaosa2012,Nagaosa2012a,Nagaosa2013,Ye1999,Bruno2004}. Similar THE signal was also observed in 45-nm-thick Bi-embedded Cr$_2$Te$_3$ samples~\cite{Note1}. The topological Hall resistivity
can be extracted from the total Hall resistivity $\rho_{yx}$ in the form
$\rho_{\text{THE}}=\rho_{yx}-R_oH-R_aM$~\cite{Note1}. Fig.~\ref{fig:figure2}~(b)
shows the $H$-dependence of $\rho_{\text{AHE}}$ and $\rho_{\text{THE}}$
at $T=35$~K, where we also plot $\rho'_{yx}\equiv\rho_{yx}-R_oH$
($=\rho_{\text{THE}}+\rho_{\text{AHE}}$) for comparison. Obviously,
$\rho_{\text{AHE}}(H)$ and $\rho'_{yx}(H)$ almost coincide with
each other, except in the region near the coercive field where the THE
appears. The resultant $\rho_{\text{THE}}$ and $\rho_{\text{AHE}}$ at
85~K are plotted in Fig.~\ref{fig:figure2}~(c). Fig. \ref{fig:figure2}~(d) is a phase diagram illustrating the strength of THE signals. Obviously, the THE occurs only in the
vicinity of the coercive field, implying that the appearance of the
non-zero spin chirality is associated with the magnetization reversal.
It is worth noting that the maximum $\rho_{\text{THE}}$ in our sample
of Bi-embedded Cr$_2$Te$_3$ is around 1300~n$\Omega\cdot$cm, which is
larger than that of
SrRuO$_3$/SrIrO$_3$~\cite{Matsuno2016} and one order of magnitude larger
than that of the B20 chiral magnet (\textit{e.g.}, MnSi)~\cite{Neubauer2009}.

\emph{Magnetic susceptibility plateau and magnetoresistivity anomaly --}
Fig.~\ref{fig:figure3} shows the temperature dependence of magnetization
\textit{M}, magnetic susceptibility $\chi(H) = \partial M(H)/\partial H$, and
magnetoresistivity $\rho_{xx}$, respectively. Right in the region near the
coercive field, the $M(H)$ curve shows an abnormal magnetization tail, which
corresponds to a striking susceptibility plateau in the $\chi(H)$ curve, as
indicated by arrows in Fig.~\ref{fig:figure3}~(a) and (b).  Furthermore, as
evident in Fig.~\ref{fig:figure3}~(c), the $\rho_{xx}(H)$ curve also exhibits a
hump feature near $H_c$, indicative of a new scattering source for electronic
transport. All these abnormal behaviors persist up to 115 K, similar to the THE
shown in Fig.~\ref{fig:figure2} (c). In the field region where THE appears,
  the $M(H)$ curve exhibits a tail, indicating the emergence of nontrivial spin
  texture. 
The coincidence of the active regions of 
THE, magnetization tail, susceptibility plateau, and magnetoresistivity hump strongly
suggests the same physical origin of them.

\emph{Discussions and conclusions --}
As a well established
theory~\cite{Nagaosa2012,Nagaosa2012a,Nagaosa2013,Ye1999,Bruno2004}, the
magnetic skyrmion can induce the THE owing to its nonvanishing spin
chirality. Skyrmions can be identified directly
by mapping the spin textures using small angle neutron scattering~\cite{Muehlbauer2009}, Lorentz
transmission electron microscopy~\cite{Yu2010},
spin-resolved scanning tunneling
microscopy~\cite{Romming2013}, resonant x-ray scattering~\cite{Langner2014}, or
even magneto-optics~\cite{Finazzi2013}. However, the direct mapping of skyrmions
is beyond the scope of this work, and is left to future investigations.
We have checked our experimental results very carefully, and found that all our experimental details can be
explained  in terms of magnetic skyrmion scenario. Furthermore, the skyrmion is
also supported in the numerical simulations with the input parameters derived
from the first-principles of our system~\footnote{L.P. Jin, X. Chen, L.
  Zhou, J. Chen, H. He, G. Wang, D.P. Yu, B. Xi, F. Ye and J.W. Mei,
  ``Topological paramagnetism in the ferromagnet embedded with
  spin-orbit-coupling nanosheets,'' to be submitted.}. 

In Cr$_{2}$Te$_{3}$ with Bi nanosheet embedded, the magnetization saturates with all the local moments fully polarized
in the high fields $H>H_c$, hence the scalar spin chirality
$\vec{S}_i\cdot(\vec{S}_j\times\vec{S}_k)$ is zero and no THE
occurs. Once the field reverses and exceeds the coercive field $H_c$,
the moments in the bulk Cr$_2$Te$_3$ are flipped first, and those near
the Bi-nanosheets are not completely flipped yet, but only twisted by
the DM interactions (resulting from the hybridization between Bi and Cr
orbitals), leading to the formation of magnetic skyrmions in the
vicinity of the coercive field. The nonzero spin chiralities of these
skyrmions in turn give rise to the THE.
This scenario is consistent with the measurements of magnetization and
longitudinal magnetoresistivity as shown momentarily. Skyrmions display the topological
paramagnetism~\cite{Note2}, \textit{i.e.}, the susceptibility plateaux and magnetization
tails, which is related to the topological stability of the magnetic
skyrmions and due to the slowing down of the
flippling speed of skyrmion moments near $H_c$~\cite{Bauer2016}. Futhermore,
the magnetic texture of skyrmions is also a new new scattering source accouting
for the magnetoresistivity hump near $H_c$ in Fig.~\ref{fig:figure3}~(c)~\cite{Bauer2016}.

To the best of our knowledge, it is the first time in the literature to observe
all these skyrmion-induced features, THE, topological paramagnetism and
magnetoresistivity anomaly, simultaneously in the same system. Therefore, it is
very likely that our system hosts the magnetic skyrmions in the THE region.  
Detailed information of the magnetic skyrmions in Bi-embedded Cr$_2$Te$_3$ can be also derived from our
experimental data~\cite{Note1}. We estimate the skyrmion size is  about 14~nm,
comparable with the lateral sizes of Bi nanosheets which may vary from
10~nm to 100~nm (see Fig.~\ref{fig:figure1}~(b)). Therefore, by controlling the
diameter of the Bi-nanosheets, a single skyrmion can be obtained and its
size may be further reduced in smaller Bi-nanosheets.

In conclusion, we have grown the bulk ferromagnet Cr$_2$Te$_3$ with the
black-phosphorous-like bismuth nanosheet embedded. It is an efficient experimental
approach to insert strong spin-orbit couplings into the traditional ferromagnet.
The nanosheet serves as a `seed' for the strong spin-orbit coupling planted into
the bulk ferromagnet, and generates the non-trivial magnetic texture, signaled by a large
topological Hall effect, topological paramagnetism and magnetoresistivity
anomaly. This method can be applied to other bulk ferromagnetic
systems, largely expanding the family of topological-Hall-effect materials.

\acknowledgments {We thank Z.Z. Du and H.Z. Lu for helpful discussions.
This work was supported by the National Natural Science
Foundation of China (No.~11574129, 11774143, 61734008, 51404293 and 11374135,
11774300), the National Key Research and Development Program of China
(No.~2016YFA0301703), the Natural Science Foundation of Guangdong Province
(No.~2015A030313840, and 2017A030313033), the State Key Laboratory of
Low-Dimensional Quantum Physics (No.~KF201602), Technology and Innovation
Commission of Shenzhen Municipality (No.~JCYJ20160531190254691,
JCYJ20160531190535310, KQJSCX20170727090712763, ZDSYS201703031659262,  and
JCYJ20170412152334605). J.W.M was partially supported by the program for Guangdong Introducing
Innovative and Entrepreneurial Teams (No. 2017ZT07C062).  J. Zhang and X. Zhang were supported by King Abdullah University of Science and Technology (KAUST), Office of Sponsored Research (OSR) under Award No: CRF-2015-2549-CRG4.}

\bibliography{THE}

\end{document}